\documentclass[a4paper]{article}
 
\pdfoutput=1

\usepackage[ruled,vlined,linesnumbered]{algorithm2e}
\usepackage{multirow} 
\usepackage{lmodern}
\usepackage{tikz}
\usepackage{balance}
\usepackage{url}
\usepackage{amssymb}
\usepackage{times}
\usepackage{todonotes}
\usepackage{authblk}
\usepackage{amsthm}
\usepackage{epstopdf}

\usepackage{pgfplots}
\pgfplotsset{filter discard warning=false}

\newcommand{\ignore}[1]{}

\begin{document}

\title{Bringing the Cloud to Rural and Remote Areas -- Cloudlet by Cloudlet}



\author[1]{Pekka Abrahamsson}
\author[2]{Sven Helmer}
\author[1]{Tosin Daniel Oyetoyan}
\author[2]{Stefan Brocanelli}
\author[2]{Filippo Cardano}
\author[2]{Daniele Gadler}
\author[2]{Daniel Morandini}
\author[2]{Alessandro Piccoli}
\author[2]{Saifur Salam}
\author[2]{Julian Sanin}
\author[2]{Alam Mahabub Shahrear}
\author[2]{Angelo Ventura}

\affil[1]{
Department of Computer and Information Science, 
Norwegian University of Science and Technology, 
7491 Trondheim, Norway}
\affil[2]{
Faculty of Computer Science,
Free University of Bozen-Bolzano,
39100 Bolzano, Italy}

\date{}

\maketitle


\begin{abstract}
Instead of relying on huge and expensive data centers for rolling out
cloud-based services to rural and remote areas, we propose a hardware platform
based on small single-board computers. The role of these micro-data centers is
twofold. On the one hand, they act as intermediaries between cloud services
and clients, improving availability in the case of network or power
outages. On the other hand, they run community-based services on local
infrastructure. We illustrate how to build such a system without incurring
high costs, high power consumption, or single points of failure. Additionally,
we opt for a system that is extendable and scalable as well as easy to deploy,
relying on an open design.
\end{abstract}

\section{Introduction}
\label{sec:intro}

Cloud computing is a disruptive technology that has already changed the way
many people live and conduct business. Clearly, individuals,
organizations, and businesses in developing countries are also adopting 
services such as reliable data storage, webmail, or online
social networks. However, according to a report by the United Nations
Conference on Trade and Development (UNCTAD), the rate at which this adoption 
takes place is much slower \cite{UNCTAD13}.

In Western countries the increasing demand by users is met by creating
ever-larger data centers and upgrading and extending high-speed communication
networks. In developing countries the picture looks different. The UNCTAD
report goes on saying that missing infrastructure is a major obstacle for the
uptake of cloud computing in these regions \cite{UNCTAD13}:

\begin{quote}
[\dots] whereas there were in 2011 more than 1000 secure data servers per million
inhabitants in high-income economies, there was only one such server per
million in [the least developed countries (LCDs)].
\end{quote}

We argue that trying to copy the infrastructure of Western countries is
neither feasible nor meets the requirements of developing countries, since setting
up and maintaining a number of large data centers can easily cost hundreds of
millions of dollars. Just using the infrastructure provided by the large
multinational corporations that create such data centers is also not an ideal
situation. Handing over data to foreign entities raises all kinds of issues
ranging from privacy and data protection (the local laws and regulations may
not match those of the hosting country) to national security and industrial
espionage.

\begin{figure*}
\begin{center}
\includegraphics[width=0.85\linewidth]{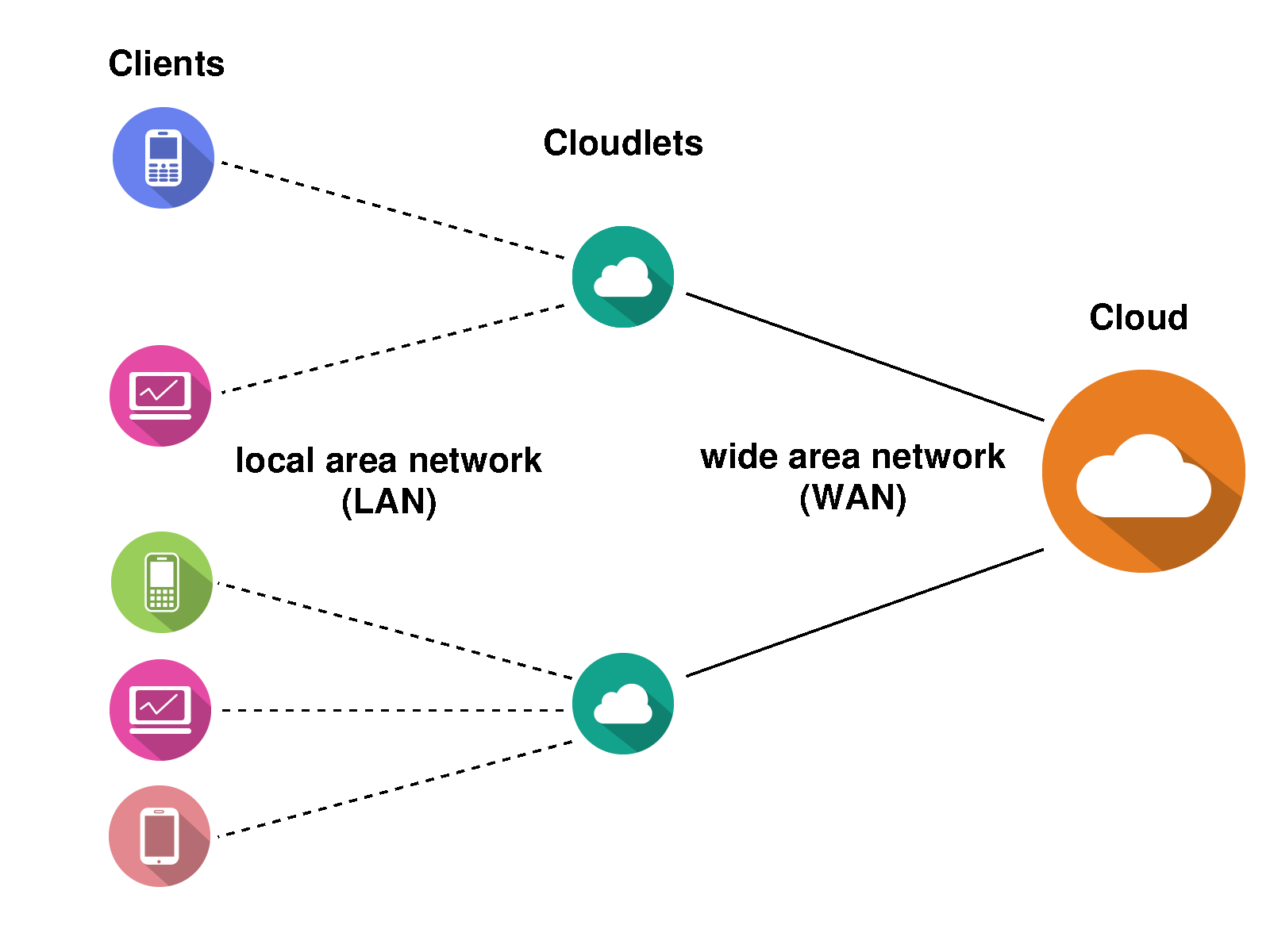}
\end{center}
\caption{Cloudlet architecture}
\label{fig:cloudlet}
\end{figure*}

Additionally, due to network outages it may be difficult to
establish reliable access to these remote servers, which is an even more acute
problem for rural areas. Then moving to the cloud to gain a reliable storage
solution would be negated by not being able to access it at all times.
Consequently, we advocate the use of small, distributed data centers that can be adapted to
the local needs in a bottom-up, community-based fashion. These can serve as
hubs and relays for (large-scale) remote cloud services or as facilities for
local community services. Pushing some of the data storage and computation to
the client side is known as a {\em cloudlet} architecture 
\cite{KLS11,SBCD09,Yang10}, 
and is done to improve the availability and durability of stored data, as
well as to lower network latencies in mobile
devices. Figure~\ref{fig:cloudlet} shows a typical cloudlet architecture: the
clients are not directly connected to the cloud but via cloudlet servers.

It is important that the design of cloudlet data
centers considers the specific requirements of developing countries. First
of all, it has to use low-cost components that are readily available. Second,
low power consumption is a crucial criterion. Due to unreliabilities in the
provision of electricity, the data center may have to be run on solar power or
batteries for considerable stretches of time. Third, we need a robust and sustainable
system than can be operated in a harsh environment (in terms of temperatures
and weather). Finally, the data center
should be based on open platforms and standards and avoid proprietary
technology as much as possible. This will make it easier to deploy, maintain,
and repair the hardware. Additionally, an open system allows users to adapt,
extend, and scale it to their particular needs. Interestingly enough, even a
big player like Facebook is
advocating the use of open hardware in their Open Compute Project initiative
\cite{SH11}.

Here we illustrate how single-board computers, such as the Raspberry Pi (RPi), 
can serve as building blocks for computing platforms that meet the requirements
described above. Originally developed to spark the interest of school children
in computer science, it has also been discovered by hobbyists worldwide, who
use it for a wide range of projects. We have been building Raspberry Pi
clusters and experimenting with them for more than two years now. A while ago
we have started using them for teaching and training purposes and even as a
testbed for research. We believe that this technology has a lot of potential
and can make an impact on the lives of many more people by serving as a
low-cost data storage and computing platform.

The remainder of the paper is structured as follows. We discuss related work
in the next section and then describe our hardware 
design in Section~\ref{sec:hardware}, followed by a brief discussion of 
software aspects in 
Section~\ref{sec:software}. Section~\ref{sec:discussion} highlights and
explains some of our design decisions in more detail and also sketches use
cases and application domains. Finally, in Section~\ref{sec:concl} we conclude
with a brief summary.

\section{Related Work}
\label{sec:related}

There are many studies investigating the adoption of cloud computing, for some
overview and survey papers, 
see \cite{DCC10,GKW12,SuKa11}.\footnote{Actually, 
two of the co-authors of this paper
were involved in a recent study \cite{PWSHA15}.} However, most of these studies
highlight the topic from the point of view of well-developed countries. While
some inhibitors, such as security and privacy concerns, hold universally 
\cite{STA14}, others are predominantly found in developing countries and play
almost no role in Western
countries \cite{Awo14,Gre10,Ksh10}. Predominant among these are unstable power grids and
inadequate internet connectivity, translating into more or less frequent
outages. This calls for a different approach to cloud computing: before cloud
services can take off, a reliable infrastructure has to be put in place.

There are numerous publications on connectivity, communications networks, and
bandwidth, see \cite{BRCK,KKR10,NOPK14,PFH04,Pob13,RuZi12} 
for a few examples. While the networking aspect is very
important, the storage and computational aspects should not be
neglected; especially since getting the infrastructure for widespread
broadband connectivity into place will still take considerable
time. Consequently, there is a need for local data centers that can bridge the
communication gaps. Tesgera et al. propose a cloudlet-based approach to tackle
network issues in emerging regions, but do so very briefly on a very abstract
level, identifying research challenges, but not proposing any implementation
\cite{TKJ14}.

We agree with Hosman and Baikie \cite{HoBa13} that the Western
``bigger is better'' approach for building large-scale data centers is bound to fail in
developing countries, due to the particular constraints. 
Furthermore, in a study Hosman identifies the main challenges faced by
hardware deployers in these regions \cite{Hos14}: energy consumption, cost,
environment-related issues, connectivity, and maintenance and support.
These are all crucial aspects we cover by relying on 
inexpensive, low-power, rugged platforms, such as micro-computers, to build
micro-data centers.\footnote{This approach is also indicated in \cite{Hos14,HoBa13}.}
There are
various groups already working on developing Raspberry Pi clusters and similar
architectures \cite{AHPN13,TWJ13}, the challenge is to adapt these designs to the
conditions found in developing countries.

\section{Hardware}
\label{sec:hardware}

In the following we give an overview of the hardware employed in our design:
in particular the casing, network architecture, electronic components, 
power supply, and cooling.

\begin{figure}
\begin{center}
\includegraphics[width=0.72\linewidth]{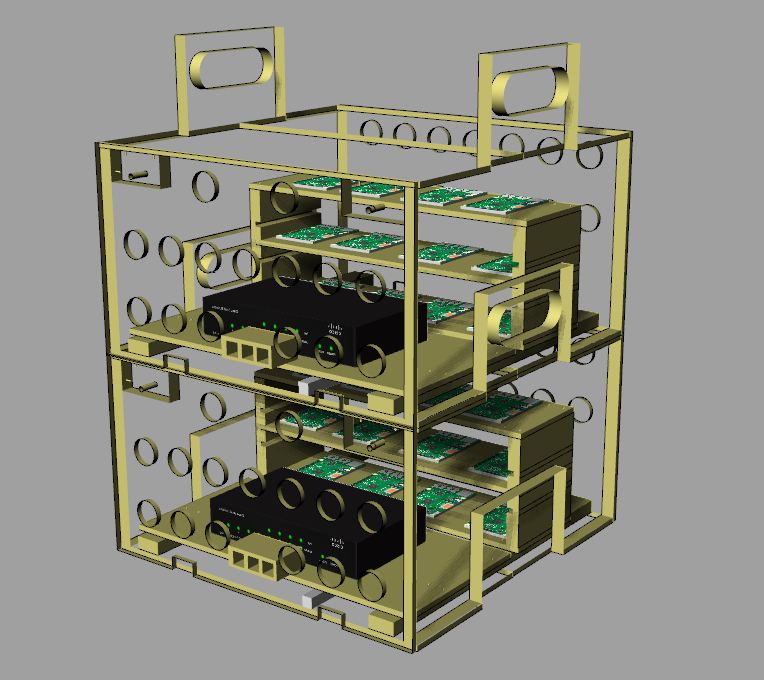}
\end{center}
\caption{Stacked boxes}
\label{fig:box}
\end{figure}

\subsection{Casing}
\label{sec:casing}

At the core of our design are modular wooden stackable boxes of size 40 x 40 x
25 cm (shown in Figure~\ref{fig:box}), offering several advantages. 
The boxes can be assembled and disassembled with a simple screw driver and for
transport the individual parts can even be stored in (hand) luggage. Once set up,
the contents of a box can be accessed without the need for any tools. The
front plate is fixed with wing screws, which can be opened with bare
hands. The boxes also feature handles, which allows their transport in
assembled state. Additionally, the electronic components are not soldered to
the casing, but fixed on shelves with screws. Consequently, the devices inside
of a box can be easily accessed, repaired, substituted, maintained, and
updated, even by a person with minimal technical skill. 
For the casing we have chosen wood, because it is an environmentally friendly
material that can be found around the world. However, this can be replaced
with material that is readily available locally.

\subsection{Network Architecture and Components}

\begin{figure}
\begin{center}
\includegraphics[width=.85\linewidth]{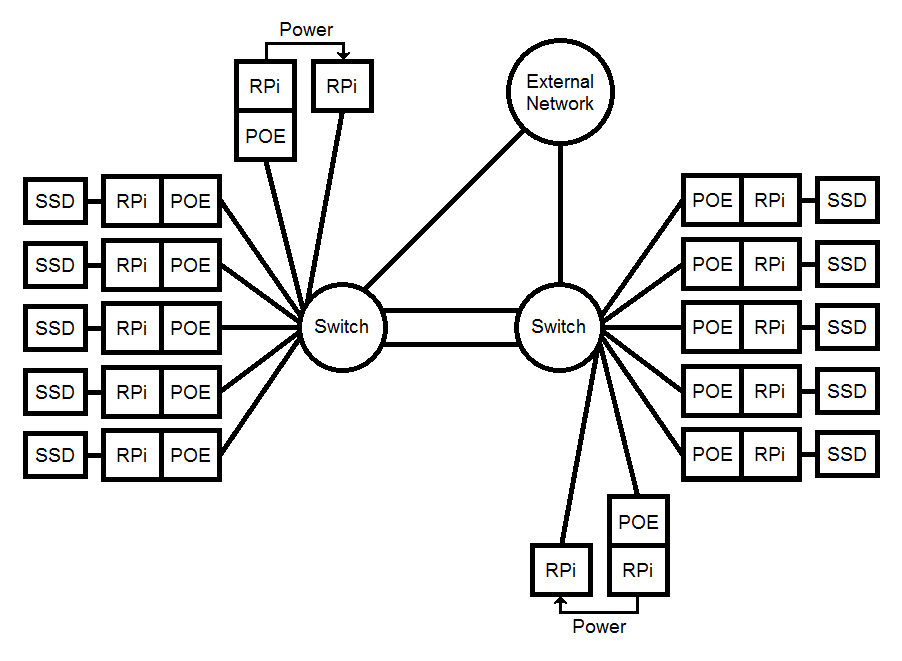}
\end{center}
\caption{Network architecture}
\label{fig:network}
\end{figure}

The default network architecture consists of two subclusters connected via two
switches, which in turn can be connected to an external
network. Figure~\ref{fig:network} shows a schematic diagram of the
architecture. (We use 8-port switches with two additional uplink ports; PoE
stands for Power over Ethernet and will be discussed later).
A subcluster, which fits into one box, comprises two control and five 
data storage/computational Raspberry
Pis, to which solid state disks are connected. 
By running two subclusters each containing multiple RPis, we avoid a single
point of failure, in case of a breakdown the data center can be kept up and
running, albeit at reduced performance. This design is also scalable: we do
not need to connect both switches to the external network, one of them can
also be connected to one or more other subclusters.

In our current design we use the new Raspberry Pi
2, model B, with a 900 MHz quad-core ARMv7 CPU and 1GByte RAM, as it exhibits
a higher performance than its predecessor, the Raspberry Pi B+. 
However, it has a slightly higher power consumption than
the B+ model, even in single-core mode. Thus, if power consumption is a
crucial issue, the Pi 2 can be swapped for the Pi B+. 
In terms of costs, the difference between the two models is about \$10, the
Raspberry Pi 2 currently sells for around \$35, the B+ model 
for \$25.\footnote{Clearly, there are additional costs for building the whole
  cluster, such as for the
  switches, solid-stage storage, batteries, cables, and so on. These very much
  depend on the size of the cluster.}
In fact, we are not
restricted to Raspberry Pi boards, other single-board computers, such as
Banana Pi, Odroid or any other device with a CPU of the ARM family would also
work. With some tinkering it would even be possible to make use of old
smart phones.

Using the power over ethernet (PoE) technology simplifies
the assembly, as we do not need separate sets of cables for communication and
power supply. Additionally, it eliminates the need for separate power supply
sockets for the Raspberry Pis, although a PoE adapter is necessary on the RPi
side. The main advantage, though, is the provision of
smart power management and monitoring techniques integrated into the
system. For example, when using the HP 2530-8-POE+ switch, PoE operation can
be disabled and enabled (at different levels of priority) for each individual
port. Clearly, this configuration has a price tag attached to it: a solution
relying on PoE will add to the costs of the system. We can make a marginal saving by
connecting the two controller RPis of a subcluster via a single PoE adapter
for supplying them with power.

\subsection{Power Supply}

The power supply, which requires a separate box, 
comprises four 12V Lithium or lead-acid batteries
configured in series with a total capacity of 200 Ah. 
The batteries, whose task it is to provide a constant source of energy, 
can be charged via solar panels or any other source of
electricity that is available. 
The lead batteries have the
advantage of being cheaper than the Lithium
ones. However, the Lithium batteries are
much better suited to higher temperatures
and will degrade much more slowly under
these conditions. The lead batteries should
not be ruled out completely, though. In a
high-altitude environment, such as the Andes
or the Himalayas, we may run into problems
with Lithium batteries and low temperatures,
as they cannot be charged at temperatures
below freezing. Thus, lead batteries would be a
better choice for these regions. The power
management is controlled by a power inverter and battery charging regulator. 
In our prototype, a 220V AC 300W sine wave power
inverter (Studer AJ-400-48) and a Lithium battery charger is used.
When using lead batteries, we can replace this with
the AJ-400-48-S model, which also integrates
a 10A 48V DC lead-acid battery charging
regulator. The AC output socket from the
power inverter can be extended by a
three-way power strip to supply a laptop or
other external diagnosis tool.

\subsection{Cooling}

An important design decision is to use only passive cooling techniques, so that
we do not incur any additional power consumption and that there are no moving
parts that can break down.
We created a set of holes on two opposite sides of the casing to make use of
the stack effect for cooling. One set of holes is found in the upper part of
one side, while the opposite set is located in the lower part. The warm air
flows out through the upper set, drawing in cooler outside air through the
lower set. For protection, the holes can be covered with anti-insect and
anti-dust nets. Depending on the placement of the boxes, additional cooling
mechanisms can be put into place, such as solar chimneys, windcatchers, or
making use of the cooler temperature underground by installing additional heat sinks.

\section{Software}
\label{sec:software}

In our group we developed tools to set up and maintain the
operating system installed on the individual nodes of the data center. We do this in an
efficient and automated manner with scripts that install a minimal version of
Debian or ArchLinuxARM on a node, register a node in the cluster, and then
update the node to fully integrate it into the cluster. We can also monitor
the activity of each node by installing our monitoring panel. 
This covers the basic infrastructure, but is not enough yet.
A cluster in which
software is deployed in a bare-metal fashion on individual nodes is not
attractive for potential users, some form of middleware is
needed. However, most off-the-shelf solutions, such as OpenStack or similar
frameworks, are usually too heavyweight for Raspberry Pi clusters.

The storage manager component of OpenStack, Swift, is relatively lightweight
compared to other components, such as Nova (computing) and Neutron
(networking). We have successfully deployed it on a Raspberry Pi cluster,
making the cluster usable as a data storage platform with data replication, 
meaning that we do not lose data in case of (partial) hardware failure. 
Currently, we are
working on extending this solution to other aspects of cloud computing.

\section{Discussion}
\label{sec:discussion}

In the following we provide more details and motivate some of the particular
design decisions we made. We also discuss application domains of our server
architecture.

\subsection{Scalability/Open Hardware and Software}

While we were designing our data center we realized that it would be crucial
to already anticipate some of the future requirements or changes that users
would make to the system to adapt it to their needs. That is the reason why we
went for a flexible design around the core of wooden boxes housing the
electronic devices. Depending on the specific
requirements in terms of power consumption, redundancy, and computational
power, a suitable number of boxes containing electronics and power supplies
can be selected and stacked on top of each other.

\subsection{Power Consumption}

In a test run, we measured the power
consumption of one box, i.e. one subcluster with seven Raspberry Pis, a
switch, and five SSDs. Generating a stress test under ArchLinux with two CPU
processes, one IO load, and one RAM load with 128MB malloc as a benchmark, one
subcluster consumed 48W running the benchmark.

We specified the capacity of the batteries assuming a load that continuously
consumes 96W (for the two subclusters). The goal was to keep the discharge
rate at 50\% for a duration of twelve hours. Regularly discharging down to low
levels has a detrimental effect on the life of the batteries. A duration of
twelve hours was chosen to be able to run the data center over night, during
which there is no sunlight for solar panels. Nevertheless, the capacity of the
batteries can be adapted to the local circumstances.

\subsection{Use Cases}

Our goal is not to compete with the high-tech environment usually found in large
cities and metropolitan areas, but to provide an alternative for more rural
areas. The proposed approach allows users to set up local servers which can be used as components
of a private or community cloud network. This is especially important when it comes to
handling sensitive data in domains such as health care and governance, as it
gives full ownership of the data and services to the persons running the data
center. Other important use cases are education and research. As we have
experienced with our own students, it is an ideal platform for teaching
practical skills in the area of distributed systems. The acquired knowledge
ranges from hardware all the way to protocols synchronizing nodes in a
network. Due to its mobility it can also be used to set up field labs for
processing data, which, for example, is generated by sensor networks
monitoring the environment. Beyond these applications, the platform may even
help in sparking entrepreneurial activity, as the initial investment is not
large and the system can be scaled out when and if the need arises. 
Moreover, upgrading the system with more efficient boards as they become
available can be done gradually, i.e., this does not have to be done in one
go, making it possible to achieve the upgrade with a number of smaller
investments rather than one big one.

\section{Conclusion and Outlook}
\label{sec:concl}

Even though cloud computing and similar services are also expanding in
developing countries, they are still far from being widely spread. 
While there are also discussions about factors inhibiting the adoption of
cloud computing in Western countries, there are factors which are unique to
the developing world, so trying to apply the Western approach to cloud
computing is very likely bound to fail. Two crucial factors, which were also
identified in a report by the World Economic Forum \cite{DuBi12}, are the lack
of both infrastructure and a skilled workforce. We believe that our micro-data
center architecture can start filling these gaps by providing an open,
inexpensive, adaptable, and extendible platform with a low power consumption,
empowering communities to take matters into their own hands. Due to these
features, it can also be rolled out in schools and universities to teach and 
grow the next generation of engineers and computer scientists.

We hope that in the near future the digital divide will not widen, as it is
currently doing in the area of cloud computing, but become narrower. However, we
believe that in order for this to happen, it is not sufficient for developing
countries to just import information technology. Ultimately, a lot of the
needed infrastructure should be developed and produced in the countries
themselves. In that light, we see our cluster as a starting point for many
more creative and innovative solutions.

\bibliographystyle{abbrv}
{\small
\bibliography{bobo}}

\begin{thebibliography}{10}

\bibitem{AHPN13}
P.~Abrahamsson, S.~Helmer, N.~Phaphoom, L.~Nicolodi, N.~Preda, L.~Miori,
  M.~Angriman, J.~Rikkil{\"a}, X.~Wang, K.~Hamily, and S.~Bugoloni.
\newblock Affordable and energy-efficient cloud computing clusters: The
  {Bolzano} {Raspberry} {Pi} cloud cluster experiment.
\newblock In {\em UsiNg and building ClOud Testbeds (UNICO) Workshop (at
  CloudCom Conf.)}, Bristol, December 2013.

\bibitem{Awo14}
R.~Awosan.
\newblock Factor analysis of the adoption of cloud computing in {Nigeria}.
\newblock {\em African Journal of Computing \& ICT}, 7(1):33--41, January 2014.

\bibitem{BRCK}
BRCK.
\newblock Modem/router specification.
\newblock \url{http://www.brck.com/specification/}, 2014.
\newblock Accessed: 2015-06-18.

\bibitem{DCC10}
T.~Dillon, C.~Wu, and E.~Chang.
\newblock Cloud computing: Issues and challenges.
\newblock In {\em Proc. 24th Int. Conf. on Adv. Inf. Networking and Appl.
  (AINA'10)}, pages 27--33, Perth, Australia, April 2010.

\bibitem{DuBi12}
S.~Dutta and B.~Bilbao-Osorio, editors.
\newblock {\em The Global Information Technology Report 2012 -- Living in a
  Hyperconnected World}.
\newblock World Economic Forum, Geneva, 2012.

\bibitem{GKW12}
G.~Garrison, S.~Kim, and R.~L. Wakefield.
\newblock Success factors for deploying cloud computing.
\newblock {\em Communications of the ACM}, 55(9):62--68, Sept. 2012.

\bibitem{Gre10}
S.~Greengard.
\newblock Cloud computing and developing nations.
\newblock {\em Communications of the ACM}, 53(5):18--20, May 2010.

\bibitem{Hos14}
L.~Hosman.
\newblock {\em Emerging Markets: Top {ICT} Hardware Challenges}.
\newblock Inveneo, San Francisco, 2014.

\bibitem{HoBa13}
L.~Hosman and B.~Baikie.
\newblock Solar-powered cloud computing datacenters.
\newblock {\em IT Professional}, 15(2):15--21, March 2013.

\bibitem{KKR10}
Y.~Kim, T.~Kelly, and S.~Raja.
\newblock {\em Building Broadband: Strategies and Policies for the Developing
  World}.
\newblock World Bank Publications, Washington, DC, 2010.

\bibitem{KLS11}
E.~Koukoumidis, D.~Lymberopoulos, K.~Strauss, J.~Liu, and D.~Burger.
\newblock Pocket cloudlets.
\newblock {\em SIGARCH Comput. Archit. News}, 39(1):171--184, Mar. 2011.

\bibitem{Ksh10}
N.~Kshetri.
\newblock Cloud computing in developing economies.
\newblock {\em IEEE Computer}, 43(10):47--55, October 2010.

\bibitem{NOPK14}
A.~Nungu, R.~Olsson, B.~Pehrson, J.~Kang, D.~Kifetew, and A.~Rustamov.
\newblock Inclusive ubiquitous access -- a status report.
\newblock In {\em 6th Int. Conf. on e‐-Infrastructure and e-‐Services for
  Developing Countries (AFRICOMM'14)}, pages 13--22, Kampala, Uganda, November
  2014.

\bibitem{PFH04}
A.~Pentland, R.~Fletcher, and A.~Hasson.
\newblock {DakNet}: rethinking connectivity in developing nations.
\newblock {\em Computer}, 37(1):78--83, 2004.

\bibitem{PWSHA15}
N.~Phaphoom, X.~Wang, S.~Samuel, S.~Helmer, and P.~Abrahamsson.
\newblock A survey study on major technical barriers affecting the decision to
  adopt cloud services.
\newblock {\em Journal of Systems and Software}, 103:167--181, May 2015.

\bibitem{Pob13}
M.~Poblet.
\newblock Affordable telecommunications: A new digital economy is calling.
\newblock {\em Australian Journal of Telecommunications and the Digital
  Economy}, 1(1):12.1--12.19, November 2013.

\bibitem{RuZi12}
S.~Ruponen and J.~Zidbeck.
\newblock Testbed for rural area networking -- first steps towards a solution.
\newblock In {\em 4th Int. Conf. on e‐-Infrastructure and e-‐Services for
  Developing Countries (AFRICOMM'14)}, pages 14--23, Yaounde, Cameroon,
  November 2012.

\bibitem{SBCD09}
M.~Satyanarayanan, P.~Bahl, R.~Caceres, and N.~Davies.
\newblock The case for {VM}-based cloudlets in mobile computing.
\newblock {\em Pervasive Computing, IEEE}, 8(4):14--23, 2009.

\bibitem{SH11}
D.~Schneider and Q.~Hardy.
\newblock Under the hood at {Google} and {Facebook}.
\newblock {\em IEEE Spectrum}, 48(6):63--67, 2011.

\bibitem{STA14}
M.~Shabalala, P.~Tarwireyi, and M.~Adigun.
\newblock Addressing privacy in cloud computing environment.
\newblock In {\em 6th Int. Conf. on e‐-Infrastructure and e-‐Services for
  Developing Countries (AFRICOMM'14)}, pages 144--153, Kampala, Uganda,
  November 2014.

\bibitem{SuKa11}
S.~Subashini and V.~Kavitha.
\newblock A survey on security issues in service delivery models of cloud
  computing.
\newblock {\em Journal of Network and Computer Applications}, 34(1):1 -- 11,
  2011.

\bibitem{TKJ14}
C.~Tesgera, M.~Klein, and A.~Juan-Verdejo.
\newblock A cloudlet-based approach to tackle network challenges in mobile
  cloud applications.
\newblock In {\em Proc. of the Int. Conf. on Advances in ICT for Emerging
  Regions (ICTer'14)}, page 253, Colombo, Sri Lanka, December 2014.

\bibitem{TWJ13}
P.~Tso, D.~White, S.~Jouet, J.~Singer, and D.~Pezaros.
\newblock The {Glasgow} {Raspberry} {Pi} cloud: A scale model for cloud
  computing infrastructures.
\newblock In {\em The 1st Int. Workshop on Resource Management of Cloud
  Computing}, Philadelphia, Pennsylvania, 2013.

\bibitem{UNCTAD13}
UNCTAD.
\newblock {\em Information Economy Report 2013. The Cloud Economy and
  Developing Countries}.
\newblock United Nations Publication, Geneva, 2013.

\bibitem{Yang10}
Z.~Yang, B.~Y. Zhao, Y.~Xing, S.~Ding, F.~Xiao, and Y.~Dai.
\newblock {AmazingStore}: available, low-cost online storage service using
  cloudlets.
\newblock In {\em Proc. of 9th Int. Workshop on Peer-to-Peer Systems
  (IPTPS'10)}, San Jose, California, 2010.

\end{thebibliography}

\end{document}